\documentstyle[twocolumn, aps]{revtex}

\begin{document}
\title{On quantum teleportation with beam-splitter-generated entanglement}
\author{Ngoc-Khanh Tran and Olivier Pfister\cite{op}} 
\address{Department of Physics, University of Virginia, 382 McCormick Rd.,
Charlottesville, VA 22904-4714}
\date{July 13, 2001} 
\wideabs{ 
\maketitle
\begin{abstract}
Following the lead of Cochrane, Milburn, and Munro \protect{[Phys.\ Rev.\ A
{\bf 62}, 062307 (2000)]}, we investigate theoretically quantum
teleportation by means of the number-sum and phase-difference variables. We
study Fock-state entanglement generated by a beam splitter and show that
two-mode Fock-state inputs can be entangled by a beam splitter into close
approximations of maximally entangled eigenstates of the phase difference
and the photon-number sum (Einstein-Podolsky-Rosen --- EPR --- states).
Such states could be experimentally feasible with on-demand single-photon
sources. We show that the teleportation fidelity can reach near unity when such
``quasi-EPR'' states are used as the quantum channel.
\end{abstract}
\pacs{PACS number(s): }
}

\section{Introduction}

Quantum teleportation~\cite{Bennett,Vaidman}, ``the disembodied transport
of an unknown quantum state from one place to another'' \cite{Furusawa}, is
a cornerstone of quantum information. It is of primordial importance for
communication between quantum computers, and the realization of quantum
gates \cite{Gottesman} and quantum error correction \cite{Vedral}. Quantum
teleportation is based on maximally entangled states, a purely quantum
mechanical feature, initially little noticed outside research on the
fundamental issues of quantum theory such as the Einstein-Podolsky-Rosen
(EPR) paradox \cite{EPR}. Producing, preserving, and detecting high quality
entanglement is an experimental challenge in making reliable quantum
teleporters. Initial experiments, based on discrete and finite Hilbert
spaces, have been successful in proving the principle but hindered by low
efficiency in the production and detection of entangled photons
\cite{Zeilinger,DeMartini,Shih}. The use of continuous quantum variables
for teleportation
\cite{Vaidman,BraunsteinKimble,VanEnk,BraunsteinMilburn,Cochrane} offers
more straightforward detection methods and also the interesting access to
an infinite Hilbert space, much richer in possibilities for encoding
quantum information. The first continuous-variable teleportation experiment
\cite{Furusawa,BraunsteinKimble} used quadrature-squeezed electromagnetic
fields and beam-splitter entanglement, and was based on the experimental
realization \cite{Ou,Leuchs} of the EPR paradox using continuous quantum
optical variables \cite{Reid,Reid2}.

Another set of interesting, in part continuous, variables is constituted by
the photon number and the phase, which are canonically conjugate in the
same sense as energy and time \cite{fn}. The use of these variables has
been proposed to realize the EPR paradox \cite{Reid} and the corresponding
maximally entangled states are therefore usable for teleportation by means
of the detection of the photon number difference and phase sum
\cite{BraunsteinMilburn}, or, more practically, of the photon number sum
and phase difference \cite{Cochrane}. In the latter reference, by Cochrane,
Milburn, and Munro (CMM), quantum teleportation using phase-difference
eigenstates as the EPR entanglement resource is proposed and studied in
detail. At the difference of CMM, we follow, in this article, the
definition of Luis and S\'anchez-Soto \cite{LuisSanchez-Soto} for
phase-difference eigenstates, or, rather, relative-phase eigenstates, and
also consider the most general ones \cite{Trifonov}. Because how to create
such a quantum state in the laboratory is not yet known for more than 2
photons, CMM also explored the use of approximate --- but much simpler and
possibly experimentally realizable using on-demand single-photon sources
--- EPR states such as created when a twin Fock state such as
$|n\rangle_{a}|n\rangle_{b}$ passes through a lossless balanced beam
splitter. In their work, CMM found that, in this case, the maximum
teleportation fidelity of an arbitrary coherent state is 50\%, due to the
fact that half the quantum amplitudes of this entangled state are null. The
gist of this paper is to show that this is not the best one can hope for
beam-splitter-generated entanglement, and we show that barely more
complicated states can in fact yield near-unity teleportation fidelities.

The outline of this paper is as follows. In Section II we derive and
evaluate close-to-maximally entangled, or quasi-EPR, states that can be
created by sending two-mode Fock states through a lossless beam splitter,
balanced or not. In Section III we use these quasi-EPR states as the
entanglement resource in the number-phase teleportation scheme, and analyze
the fidelity.

\section{Generation of approximate EPR phase-difference eigenstates by a
beam splitter}
 
\subsection{The Schwinger representation}

We begin by recalling the definition of the Schwinger representation
\cite{Schwinger}, widely used in quantum optics, of a nondegenerate
two-mode field in terms of a fictitious spin. This spin is defined as
\begin{eqnarray}  \label{j}
\bbox{J}= \left(\begin{array}{c} J_x \\ J_y \\ J_z
\end{array}\right)={\frac{1}{2}} \left(\begin{array}{c} a^\dagger b +
b^\dagger a \\ -i(a^\dagger b - b^\dagger a) \\
a^\dagger a - b^\dagger b \end{array}\right)\\
\label{modj2}
\bbox{J}^2\ =\ \frac{a^\dagger a+b^\dagger b}{2} \left(\frac{a^\dagger
a+b^\dagger b}{2}+1\right),
\end{eqnarray}
where $a$ and $b$ are the photon anihilation operators of each mode. The
physical meaning of these operators is the following
\cite{LuisSanchez-Soto,Kim}: $\bbox{J}^{2}$ represents the total photon
number, $J_{z}$ the photon number difference between the two modes, and
$J_{x,y}$ are the phase-difference, or interference, quadratures. It stems
from this that $e^{i\theta J_{z}}$ is the relative phase shift operator and
$e^{i\theta J_{x}}$ is the rotation carried out by a beam splitter
(homo-/heterodyne measurements). The eigenstates of the fictitious spin are
the two-mode Fock states
\begin{equation}  \label{es}
|j\,m\rangle_z=|n_{a}\rangle_a |n_{b}\rangle_b,
\end{equation}
and the respective eigenvalues of $J^{2}$ and $J_{z}$ are given by
\begin{eqnarray} 
j & =& \frac{n_a+n_b}{2} = \frac{N}{2}, \\
\label{qn}
m & = & \frac{n_a-n_b}{2}.
\end{eqnarray}
The Schwinger representation thus makes use of the homomorphism from SU(2)
onto SO(3), which allows one to represent any unitary operation on the
two-mode field by a rotation. The general SU(2) transformation
\begin{equation}  \label{su2}
\left(\begin{array}{c} a_{1} \\ b_{1} \end{array}\right)
= \left(\begin{array}{cc}
\cos\frac{\beta}{2} e^{\frac{i}{2}(\alpha+\gamma)} &
\sin\frac{\beta}{2} e^{\frac{i}{2}(\alpha-\gamma)} \\
-\sin\frac{\beta}{2} e^{-\frac{i}{2}(\alpha-\gamma)} &
\cos\frac{\beta}{2} e^{-\frac{i}{2}(\alpha+\gamma)}
\end{array}\right)
\left(\begin{array}{c} a_{0} \\ b_{0} \end{array}\right),
\end{equation}
where $\alpha$, $\beta$, and $\gamma$ are the Euler angles, corresponds to
the rotation operator $e^{-i\alpha J_z}\, e^{-i\beta J_y}\,e^{-i\gamma
J_z}$ ($\hbar=1$). A lossless beam splitter corresponds to the values
\begin{equation}\label{beta}
\alpha=-\gamma=\pi/2,\quad \beta=\pm 2\arccos\sqrt{R},
\end{equation}
where $R$ is the reflectivity of the beam splitter (the transmittivity $T$
is such that $R+T=1$).

\subsection{Ideal number-phase EPR states}

By definition, a maximally entangled two-mode state, or EPR state, is a
two-mode quantum state
\begin{equation}\label{2mod}
|EPR\rangle\ =\ \sum_{k,l}\,s_{kl}\,|k\rangle_{a}|l\rangle_{b}
\end{equation}
such that any reduced (single-mode) density matrix of this state ${\rm
Tr}_{a,b}(|EPR\rangle\langle EPR|)$ is proportional to the identity matrix,
which yields
\begin{equation}\label{maxent}  
\sum_{k}\,s_{kl}s^{*}_{kl'}=\delta_{ll'},
\quad\sum_{l}\,s_{kl}s^{*}_{k'l}=\delta_{kk'}.
\end{equation}
An example of EPR state is the eigenstate, introduced by Luis and
S\'anchez-Soto \cite{LuisSanchez-Soto}, of the operators of the
photon-number sum and phase difference of two modes $a$ and $b$. Heeding
the point made by Trifonov {\em et al.\ } \cite{Trifonov}, we will call
this state a {\em relative-phase} eigenstate rather than a phase-difference
eigenstate, thus recalling that the formal definition of a two-mode quantum
phase difference operator does not coincide with the (problematic)
definition of two single-mode quantum phase operators \cite{QuantumPhase}.
The relative-phase eigenstate is
\begin{equation}\label{EPR2}
|\phi^{(N)}_{r}\rangle\ =\ \frac{1}{\sqrt{N+1}}\sum_{n=0}^{N}\,e^{i
n\phi^{(N)}_{r}} \,|n\rangle_{a}|N-n\rangle_{b},
\end{equation}
where $\phi^{(N)}_{r}=\phi_{0}+2\pi r/(N+1)$ is the phase difference, $N$
is the total photon number, $\phi_{0}$ an arbitrary phase origin, and
$r\in[0,N]$. The phase difference is adequately defined with resolution
1/$N$, i.e.\ at the Heisenberg limit. In the Schwinger representation,
Eq.~\ref{EPR2} becomes
\begin{equation}\label{EPRSchw}
|\phi^{(2j)}_{r}\rangle\ =\ \frac{1}{\sqrt{2j+1}}\ \sum_{m=-j}^{j}\,e^{i
m\phi^{(2j)}_{r}}\,|j m\rangle_{z}.
\end{equation}
It is worth noting that the state
\begin{equation}\label{weird}
|\{\theta^{(N)}\}\rangle\ =\
\frac{1}{\sqrt{N+1}}\sum_{n=0}^{N}\,e^{i\theta^{(N)}_{n}}\,|n\rangle_{a}|N-n\rangle_{b}
\end{equation}
can always be considered a relative-phase eigenstate, however involved or
even arbitrary the real set $\{\theta^{(N)}_{n}\}_{n}$ may be with respect
to $n$ \cite{Trifonov}. The whole basis can always be constructed by
applying the $N+1$ phase shift operators $e^{i\phi^{(N)}_{r}J_{z}}$ to
$|\{\theta^{(N)}\}\rangle$. However, successful quantum teleportation will
demand full initial knowledge of $\{\theta^{(N)}_{n}\}_{n}$, which is the
relative phase of the entanglement between Alice and Bob
\cite{VanEnkPhase}, and we will show later that the number-phase
teleportation protocol becomes complicated if $\theta^{(N)}_{n}$ is not
linear in $n$.

Finally, let us recall that maximal entanglement is only really attained
when $N\rightarrow\infty$.

\subsection{Generation of EPR and quasi-EPR states}

The experimental realization of relative-phase eigenstates is an arduous
problem. Recently, Trifonov {\em et al.\ } reported the experimental
realization of a relative-phase eigenstate (\ref{weird}) for $N=2$
\cite{Trifonov}. Their astute method uses a nonbalanced beamsplitter to
create a two-mode state, all of whose amplitudes have equal modulus. This
method is not general in the sense that it cannot work perfectly for $N>2$,
as we show below. However, the use of a beam splitter to generate EPR or
quasi-EPR states stems from quite general arguments indeed.

The studies of Heisenberg-limited interferometry
\cite{HLI,Yurke,Holland,Hillery,Kim} have led to a thorough understanding
of the subtle quantum optical properties of the beam splitter
\cite{Campos,Yurke,Holland,Hillery,Kim}. In particular, a balanced beam
splitter swaps the amplitudes and phase properties of the impinging
two-mode wave, and can also be used to entangle nonclassical optical fields
\cite{Furusawa,Leuchs2,Buzek}. As is readily seen in the Schwinger
representation, a balanced beam splitter corresponds to a $\pi/2$ rotation
around $X$ and therefore transforms a state from axis $Z$ (intensity
difference) to axis $Y$ (phase difference).

In the case of an EPR state such as Eq.~(\ref{EPRSchw}), the phase
difference is squeezed and the intensity difference is anti-squeezed.
Experimentally, this is achievable by sending an
intensity-difference-squeezed state through a beam splitter \cite{Leuchs}.
To illustrate this, let us examine the beam-splitter output of the relative
phase state (\ref{EPRSchw}):
\begin{eqnarray}\label{TrueState}
|\phi^{(2j)}_{0}\,(\beta)\rangle & = & e^{-i\beta J_{x}}\
|\phi^{(2j)}_{0}\rangle\\
& = & \frac{1}{\sqrt{2j+1}}\
\sum_{m'=-j}^{j}\,i^{-m'}\,f^{j}_{m'}\,(\beta)\,|j\,m'\rangle_{z},
\end{eqnarray}
where $d^j_{m' m}\,(\beta)=\langle j m'|e^{-i\beta J_y}|j\,m\rangle$ is a
rotation matrix element taken real by convention and proportional to a
Jacobi polynomial \cite{bie}, and where
$f^{j}_{m'}\,(\beta)=\sum_{m=-j}^{j}\,e^{i
m(\phi^{(2j)}_{0}+\pi/2)}\,d^{j}_{m' m}\,(\beta)$. This state is displayed
for $\phi_{0}=0$ in Fig.~\ref{EPRBS}. As expected, the result is a narrow
state in photon number (i.e.\ $f^{j}_{m'}\rightarrow 0$ very fast as
$|m|\rightarrow j$). 

It is straightforward to see that sending $|\phi^{(2j)}_{0}(\pi/2)\rangle$
through another, balanced ($\beta=\pm\pi/2$), beam splitter will
reconstruct the initial relative phase eigenstate
$|\phi^{(2j)}_{0}\rangle$. (One can see that $\beta\rightarrow\pi/2$ is
necessary to maximal entanglement, since the spread of the state must cover
all values of the projections of the spin.) Now, since
$|\phi^{(2j)}_{0}(\pi/2)\rangle$ contains but a few nonzero amplitudes, a
reasonable method for generating close approximations of EPR states with a
balanced beam splitter is to consider ``quantum filtered'' input states,
which are derived from $|\phi^{(2j)}_{0}(\pi/2)\rangle$ by keeping only the
lowest values of $m$ (this could be viewed as a sort of quantum Abbe
experiment, with a low-$m$-pass filter). These states are, for $N=2j$ even,
\begin{eqnarray}\label{j0}
&|j\,0\rangle_{z},\\\label{3pt}
&\left(f^{j}_{0}|j\,0\rangle_{z}+f^{j}_{1}|j\,1\rangle_{z}
+f^{j}_{-1}|j\,-1\rangle_{z}\right)/C_{1},
\end{eqnarray}
and so on, with $C_{\mu}=(\sum_{m=-\mu}^{\mu}|f^{j}_{m}|^{2})^{1/2}$, and
are
\begin{eqnarray}\label{2pt}
&\left(|j\,\scriptstyle{\frac{1}{2}}\displaystyle\rangle_{z}+|j
\scriptstyle{-\frac{1}{2}}
\displaystyle\rangle_{z}\right)/\sqrt{2},\\\label{4pt}
&\left(f^{j}_{\frac{1}{2}}
|j\,\scriptstyle{\frac{1}{2}}\displaystyle\rangle_{z}
+f^{j}_{-\frac{1}{2}}|j \scriptstyle{-\frac{1}{2}}\displaystyle\rangle_{z}
+f^{j}_{\frac{3}{2}}|j \scriptstyle{\frac{3}{2}}\displaystyle\rangle_{z}
+f^{j}_{-\frac{3}{2}}|j
\scriptstyle{-\frac{3}{2}}\displaystyle\rangle_{z}\right)/C_{\frac{3}{2}},
\end{eqnarray}
and so on, for $N=2j$ odd. Sending these states through a beam splitter
gives output states closely resembling EPR states, as we will now see. We
call these output states quasi-EPR states.

We start with the simplest one (\ref{j0}), which has already been
considered by CMM. We define the general state rotated by a beam splitter
as
\begin{eqnarray}\label{rot}
|j\,m \,(\beta)\rangle & = & e^{-i\beta J_{x}}\ |j\,m\rangle_{z}\\
& = & \sum_{m'=-j}^{j}\ i^{m-m'}\ d^{j}_{m' m}\,(\beta)\ |j\,m'\rangle_{z}.
\end{eqnarray}
Figure \ref{3DNNBS} displays the modulus of the quantum amplitudes of
$|j\,0\,(\beta)\rangle$, versus $\beta$ (\ref{beta}) and $m$. 

One can see that $\beta\rightarrow\pi/2$ is still necessary for maximal
entanglement. The very value $\beta=\pi/2$ leads to a problem, however,
because every other amplitude of the state is zero \cite{Campos,Kim}. This
was recalled by CMM when they investigated the use of this state as a
teleportation channel and found that, because of this, teleportation
fidelity was bounded by 50\% (to the notable exception of some
Schr\"odinger cat states). This situation, however, is changed if one
considers an ever-so-slightly imbalanced beam splitter: Fig.~\ref{NN855BS}
dispays the modulus and phase of the coefficients of $|10\ 0(90^{\rm
o})\rangle$ and $|10\ 0(85.5^{\rm o})\rangle$. 

Clearly, $|10\ 0(85.5^{\rm o})\rangle$ is closer to an EPR state than $|10\
0(90^{\rm o})\rangle$: it has the same spread but much more even amplitudes
and no zeroes at all. The phase distribution is not constant but this just
means that it is a general relative phase state of the form of
Eq.~(\ref{weird}), which is still a legitimate EPR state. In fact, $|10\
0(85.5^{\rm o})\rangle$ is the best quasi-EPR state for all $\beta$. In
general, we find that the angle $\beta_{Q}$ that gives the best quasi-EPR
state is given by the following phenomenological formula:
\begin{equation} \label{lim} 
\beta_{Q}\ =\ \frac{\pi}{2}\left(1-\frac{1}{N}\right).
\end{equation}
To test Eq.~(\ref{lim}), we have plotted on Fig.~\ref{NN895BS} the state
$|j\,0\,(\beta_{Q})\rangle$ for $j$ = 100, 1000, and 10000. 

Note that all digits in the angle in Fig.~\ref{NN895BS} are significant,
which points to an interesting situation. Let us assume that on-demand
single-photon sources become a reality (not an unreasonable asumption),
which would allow the production of $|j\,0\rangle_{z}$ in the laboratory.
Equation (\ref{lim}) nevertheless poses a serious experimental constraint
on the tolerance of the beam splitter reflectivity $R$, because the
required precision on $\beta$, i.e.\ on $R$, increases with $N$ if one
wants to resolve $|j\,0\,(\beta_{Q})\rangle$ from $|j 0(\pi/2)\rangle$ and
its numerous inconvenient zeroes. Roughly, $\Delta
R\sim\Delta\beta\sim1/N$. Since a beam splitter using state-of-the-art
optical coatings and polishing cannot give more than the (already
irrealistic) $\Delta R\sim 10^{-6}$, hence $N$ cannot exceed $10^{6}$
photons.

In fact, by taking a closer look at Fig.~\ref{3DNNBS}, one can see that the
amplitudes present $1/N$-period oscillations versus $\beta$. These
oscillations are of significant contrast, with the state amplitudes often
reaching zero. This poses a problem for experimentally defining
$|j\,0\,(\beta)\rangle$ as $j$ increases.

This problem disappears, however, as soon as one uses a more elaborate
input state, such as (\ref{2pt}). Such an input state could be obtained
using stimulated emission from a single atom, starting from a $|j
0\rangle_{z}$ state and having the two beams shine simultaneously and
noncollinearly on the excited atom. One will also want to have fast
nonradiative decay from the ground state of the transition so as to prevent
subsequent absorption. Figure \ref{3D2pt} displays the state
$(|j\,\frac{1}{2}\,(\beta)\rangle
+|j\,-\frac{1}{2}\,(\beta)\rangle)/\sqrt{2}$. 

Even though the $1/N$ oscillations are still present, they are
significantly attenuated as there are no zero amplitudes, even at
$\beta=\pi/2$. One can therefore use $\beta=\pi/2$ in this case, which
presents the nonegligible advantage of yielding a constant phase
distribution for this state $|j\,\frac{1}{2}(\pi/2)\rangle+|j
-\frac{1}{2}(\pi/2)\rangle$ (unlike in Fig.~\ref{NN855BS}). This is of
importance for the teleportation protocol, as we will see in the next
Section. Finally, it is straightforward and unsurprising to show that the
more elaborate (less low-$m$-pass filtered) reconstructions (\ref{3pt}) and
(\ref{4pt}) give even better results: more even amplitudes at still
constant phase. We will, however, restrain our investigations to the two
states $|j\,0\,(\beta_{Q})\rangle$ and
$|j\,\frac{1}{2}(\pi/2)\rangle+|j\,-\frac{1}{2}(\pi/2)\rangle$, which are
the simplest ones and are also within reasonable reach of foreseeable
future technology.

\section{Number-phase quantum teleportation}

In this section we briefly recall the definition of number-phase
teleportation \cite{Cochrane} and extend it to general relative-phase
states. We then consider the use of the quasi-EPR states derived above.

\subsection{Ideal entanglement resource}

Quantum teleportation relies on a maximally entangled state shared by the
sender, Alice, and the receiver, Bob. The entanglement concerns two
physical systems, $\cal A$ and $\cal B$ respectively. Alice is in
possession of $\cal A$ and also of system $\cal T$, whose ``target'' state
is the quantum information she needs to transmit. The teleportation process
consists, for Alice, in making a joint measurement on $\cal T$ and $\cal A$
such that both are projected onto a maximally entangled state. This
prevents Alice from obtaining any quantum information about the target
state, which as such is destroyed in the process. In turn, said target
state is transferred, by ``entanglement transitivity'' from $\cal T$ to
$\cal B$, i.e.\ to Bob, who may then reconstruct the exact target state on
$\cal B$ using the classically transmitted results of Alice's measurements
(which contain no quantum information whatsoever). The conceptually
difficult part is to figure out what measurements should be used by Alice
to maximally entangle her two systems $\cal A$ and $\cal T$. This question
was answered by Vaidman in connection with the EPR paradox \cite{Vaidman}.

In the case of number-phase teleportation, use is made of the commuting
operators number sum $\hat N_{T}+\hat N_{A}$ and (Hermitian) relative phase
\cite{LuisSanchez-Soto}
\begin{equation}\label{phaseop}
\hat\phi_{TA}\ =\ \sum_{N=0}^{\infty}\sum_{r=0}^{N+1}\,|\phi^{{(N)}}_{r}\rangle
\phi^{{(N)}}_{r}\langle\phi^{{(N)}}_{r}|,
\end{equation}
whose measurements project the joint $\cal T$-$\cal A$ state onto a joint
eigenstate of the total number and the relative phase such as
Eq.~(\ref{EPR2}). If the same type of entangled state is shared between
Alice and Bob, perfect teleportation can in principle be achieved. Let us
consider the general case where the initial total state is
\begin{equation}  
\begin{array}{l}
|\psi\rangle_{T}\otimes|\phi^{(N)}_{r}\rangle_{AB}\ \equiv\\ \\  
\displaystyle\frac{1}{\sqrt{N+1}}\ \sum_{m=0}^{\infty}\,c_{m}\,|m\rangle_{T}
\ \sum_{n=0}^{N}\,e^{i n\phi^{(N)}_{r}}
\,|n\rangle_{A}|N-n\rangle_{B}.
\end{array}  
\end{equation}
We assume Alice's measurements yield the eigenvalues $\langle \hat
N_{T}+\hat N_{A}\rangle=q$ and
$\langle\hat\phi_{TA}\rangle=\phi^{(q)}_{s}$. The joint TA state is thus
left in $|\phi^{(q)}_{s}\rangle_{TA}$, and the total state after Alice's
measurement is
\begin{eqnarray}  
&\displaystyle|\psi_{M}\rangle \ =\ e^{iq\phi^{(N)}_{r}}\
|\phi^{(q)}_{s}\rangle_{TA}\otimes |\psi_B\rangle_{B}\\
&\displaystyle|\psi_B\rangle_{B}\ =\ C(q)\
\sum^{q}_{k_{0}}\,e^{-ik(\phi^{(N)}_{r} + \phi^{(q)}_{s})} \,c_{k}
\,|k+N-q\rangle_{B},
\end{eqnarray}
where $k_{0}={\rm Max}[0,q-N]$ and
$C(q)=(\sum_{k=k_{0}}^{q}\,|c_{k}|^{2})^{-1/2}$. To exactly recover the
target state, Bob must then perform, on $\cal B$, a photon number shift
\cite{Cochrane} of $q-N$ and a phase shift of
$\phi^{(N)}_{r}+\phi^{(q)}_{s}$ i.e.
\begin{equation}  
|\psi_{out}\rangle_{B}\ =\ e^{i(\phi^{(N)}_{r} + \phi^{(q)}_{s})J_{z}}\
{\cal P}_{q-N}\ |\psi_B\rangle_{B}.
\end{equation}
(In the particular case where $q=N$, $\phi^{(q)}_{s}=-\phi^{(N)}_{r}$, Bob
does not have to do anything). One can see from this that the phase
distribution of the initial entanglement resource has to be corrected for,
along with Alice's measurement result. If this distribution is unknown or
too complicated to correct, teleportation will fail. This correction can
also be made by Alice, by simply shifting her phase operator, i.e.\ using
\begin{equation}\label{phaseop2}
e^{-i\phi^{(N)}_{r}J_{z}}\hat\phi_{TA}e^{i\phi^{(N)}_{r}J_{z}}
\end{equation}
instead of $\hat\phi_{TA}$. [$J_{z}=(\hat N_{T}-\hat N_{A})/2$ here.]

Note that this requirement that the entanglement phase be perfectly known
is a very general one and not at all specific to our particular choice of
the optical phase variable for the teleportation protocol. This was pointed
out by van Enk in Ref.~\cite{VanEnkPhase}.

In light of what precedes, it is interesting to investigate the use of the
general relative phase state $|\{\theta^{(N)}\}\rangle$ given by
Eq.~\ref{weird} (which still is a perfect EPR state). Indeed, if the phase
distribution is more complicated than the mere relative phase offsets of
Eq.~(\ref{EPR2}), Bob will be faced with problems reconstructing the target
state. If the initial entanglement is given by Eq.~\ref{weird},
\begin{equation}  
|\psi\rangle_{T}\otimes|\{\theta^{(N)}\}\rangle_{AB},
\end{equation}
and the phase difference operator by Eq.~(\ref{phaseop}), then the
post-measurement total state is
\begin{eqnarray}  
&\displaystyle|\psi_{M}\rangle \ = \ e^{i\frac{q}{2}\phi^{(q)}_{s}}\
|\phi^{(q)}_{s}\rangle_{TA}\otimes |\psi_B\rangle_{B}\\
&\displaystyle|\psi_B\rangle_{B} = C(q) \sum^{q}_{k=k_{o}}
e^{i\theta^{(N)}_{q-k}} e^{-ik\phi^{(q)}_{s}} c_{k} |k+N-q\rangle_{B},
\end{eqnarray}
which shows that Bob needs more than a phase shift to properly reconstruct
$|\psi\rangle$, even if $\{\theta^{(N)}_{n}\}_{n}$ is fully known. What is
needed is the unitary transformation $U_{\{\theta^{(N)}\}}$ that transforms
$|\{\theta^{(N)}\}\rangle$ into a ``flat-phase'' state for which
$\theta^{(N)}_{n}=\phi^{(N)}_{0}=$ cst, $\forall n$. This transformation
may be applied by Bob, as $U_{\{\theta^{(N)}\}}|\psi_B\rangle_{B} =
|\psi_{out}\rangle_{B}$, or by Alice, by measuring
$U_{\{\theta^{(N)}\}}\hat\phi_{TA}U^{\dagger}_{\{\theta^{(N)}\}}$.

It is however somewhat puzzling that this additional step is needed if
$|\{\theta^{(N)}\}\rangle$ may indeed be considered as a legitimate
relative phase eigenstate, since it is used in the corresponding relative
phase measurement. There seems therefore to be a need for an {\em absolute}
phase reference in number-phase teleportation, if Alice uses a relative
phase operator. In other words yet, even though the whole relative-phase
eigenbasis --- and hence the operator --- may be generally defined based
upon any general state $|\{\theta^{(N)}\}\rangle$ with arbitrary phase
distribution $\{\theta^{(N)}_{n}\}_{n}$, it does in fact matter for
teleportation that $U_{\{\theta^{(N)}\}}$ correspond to a feasible physical
measurement, thereby limiting the generality of relative phase states
usable for teleportation. If $U_{\{\theta^{(N)}\}}$ is a phase shift, then
$\{\theta^{(N)}_{n}\}_{n}$ can only be linear in $n$ at most.

One example of such a complicated situation is the quasi-EPR state 
$|j\,0\,(\beta_{Q})\rangle$, which has the phase distribution depicted in 
Fig.~\ref{NN855BS}.

\subsection{Teleportation of a coherent state with a quasi-EPR resource} 

We now turn to the use of quasi-EPR states as the entanglement resource,
and show how an arbitrary coherent state can be successfully teleported.
Our evaluation of teleportation performance is based on the pure-state
fidelity
\begin{equation}\label{fid}  
F\ =\ |\langle\psi_{out}|\psi\rangle|^{2}.
\end{equation}
The entanglement resource is now a quasi-EPR state, i.e.\ of the general
form (\ref{2mod}), but where Eqs.~(\ref{maxent}) are not verified any more.
The entangled state is of the form (\ref{EPR2}), but with nonequal
amplitude moduli: we write
\begin{equation}\label{QEPR}  
|QEPR\rangle_{AB}\ =\ \sum_{n=0}^{N}\,s^{N}_{n}
\,|n\rangle_{A}|N-n\rangle_{B},
\end{equation}
where $\sum_{n=0}^{N}|s^{N}_{n}|^{2}=1$. As announced before, we only treat
the cases of $|j\,0\,(\beta_{Q})\rangle$ and $|j
\frac{1}{2}(\frac{\pi}{2})\rangle+|j\,-\frac{1}{2}(\frac{\pi}{2})\rangle$.
their respective decompositions in terms of Eq.~(\ref{QEPR}) are found
using Eqs.~(\ref{es}-\ref{qn},\ref{rot}), and their amplitudes are,
respectively,
\begin{eqnarray}
s^{N}_{n} & = &
i^{-n+\frac{N}{2}}\,d^{\frac{N}{2}}_{n-\frac{N}{2},\,0}(\beta_{Q}),\\
s^{N}_{n} & = & \frac{1}{\sqrt{2}}\,i^{-n+\frac{N}{2}}\,\left[
d^{\frac{N}{2}}_{n-\frac{N}{2},\,\frac{1}{2}}(\frac{\pi}{2}) - i
d^{\frac{N}{2}}_{n-\frac{N}{2},\,-\frac{1}{2}}(\frac{\pi}{2})\right].
\end{eqnarray}
The post-measurement total state is
\begin{eqnarray}  
&\displaystyle|\psi_{M}\rangle \ = \ |\phi^{(q)}_{s}\rangle_{TA}\otimes
|\psi_B\rangle_{B} \\
&\displaystyle|\psi_B\rangle_{B}\ =\ C(q)\ \sum^{q}_{k=k_{0}}\,e^{-i
k\phi^{(q)}_{s}}\, c_{k}\,s^{N}_{q-k}\,|k+N-q\rangle_{B},
\end{eqnarray}
where $C(q)=(\sum_{k=k_{0}}^{q}\,|c_{k}|^{2}|s^{N}_{n}|^{2})^{-1/2}$.
This yields
\begin{equation}
|\psi_{out}\rangle_{B}\ =\ C(q)\ \sum^{q}_{k=k_{0}}\,
c_{k}\,s^{N}_{q-k}\,|k\rangle_{B},
\end{equation}
and the fidelity 
\begin{equation}\label{fq}
F(q) \ =\ \frac{\displaystyle\left|\sum_{k=k_{0}}^{q} |c_{k}|^{2}
s^{N}_{q-k}\right|^{2}}{\displaystyle\sum_{k=k_{0}}^{q} |c_{k}|^{2}
|s^{N}_{q-k}|^{2}}.
\end{equation}
Before we plot $F(q)$ for the two quasi-EPR states, we must address the
dependence of the fidelity on the measured value $q$ of the number sum:
this implies conditional teleportation even for an ideal relative-phase
state ($s^{N}_{n} =$ cst, $\forall N, n$), which should not be.

Equation (\ref{fq}) has an upper bound, which can be found by using the
Cauchy-Schwartz inequality:
\begin{equation}
\left|\sum_{k=k_{0}}^{q} |c_{k}|^{2} s^{N}_{q-k}\right|^{2}\ \leq \
\sum_{k=k_{0}}^{q} |c_{k}|^{2} \sum_{k=k_{0}}^{q} |c_{k}|^{2}
|s^{N}_{q-k}|^{2}.
\end{equation}
The fidelity is thus bound by
\begin{equation}\label{limfq}
F(q) \ \leq\ \sum_{k=k_{0}}^{q} |c_{k}|^{2},
\end{equation}
which corresponds to the fidelity of the ideal EPR resource. The
Cauchy-Schwartz inequality thus has a precise physical meaning in this
case. Note, however, that this limit (\ref{limfq}) can be $\ll$ 1 unless
$k_{0}=0$ (i.e.\ $q\leq N$) and $q$ is very large compared to the spread of
the target state (i.e.\ $k_{max}$ such as $c_{k}\neq 0$). This can only be
achieved if $N\rightarrow\infty$, which is the rigorous condition for which
a relative-phase eigenstate is truly maximally entangled. When this is
fulfilled, the probability that $q<k_{max}$ becomes negligible and the
teleportation becomes unconditional $F(q)\leq 1$. 

Unfortunately, $N\rightarrow\infty$ cannot be satisfactorily approximated
in numerical simulations with $N\sim 10$ to 100, therefore the fidelity
displayed in our figures has high and low regions, depending on the value
of $q$. It is simple to see that, for a coherent target state
$|\alpha\rangle$ ($\alpha$ real), the high-fidelity region is given by
$q\in[k_{max},N-k_{min}]$ where $N\gg k_{max}$ i.e.\
$q\in[\alpha^{2}+\alpha,N-\alpha^{2}+\alpha]$.

Figure \ref{F3D2pt} displays the fidelity versus $q$ and the beam splitter
angle $\beta$ for the teleportation of a coherent state $\alpha=3$ with the
quasi-EPR resource $(|j\,1/2\,(\beta)\rangle +
|j\,-1/2\,(\beta)\rangle)/\sqrt{2}$.

Once again, the limited spread in $q$ of the high fidelity region is only 
due to the limitations of our computation. What is essential is that 
$F>80\%$ for $\beta=\pi/2$, thereby confirming our analysis of quasi-EPR 
states.

A more exotic illustration is the use of $|j\,0\,(\beta_{Q})\rangle$ as the
quasi-EPR resource. Because the phase distribution is not simple
(Fig.~\ref{NN855BS}), either Bob or Alice must perform a $U_{\{\theta\}}$
besides the required teleportation protocol. In this case, one possible
operation (whose physical meaning is not so clear) would be
\begin{equation}
U_{\{\theta\}}\ =\ e^{i{\rm f}(q)\frac{\pi}{2}\hat{N_{B}}^{2}},
\end{equation}
where $\hat{N_{B}}$ is the number operator for Bob's field, and 
$ f(q)=(-1)^{q} $. Were Alice and 
Bob able to perform this transformation, the fidelity would thus be as 
displayed on Fig.~\ref{Fj0}.

The peak value is $99.3\%$ at $\beta= 85.5^{\rm o}=\beta_{Q}$ ($0.5^{\rm
o}$ steps). Note that $\beta= 90^{\rm o}$ gives $F=49.8\%$, consistently
with CMM.

\section{Conclusion}

We hope to have convincingly demonstrated that efficient quantum 
teleportation is indeed possible with only approximately entangled states 
(close enough to ideal, nevertheless). We chose the simplest quasi-EPR 
states, which could be good candidates for experimentation with on-demand 
single-photon sources. Finally, we investigated the use of generalized 
relative-phase states for quantum teleportation and showed that these 
states, although usable, lead to serious complications if the basis
``generator'' $|\{\theta\}\rangle$ has a more complicated phase 
distribution than a simple phase offset.

\begin{figure}
\caption{Modulus of the quantum amplitudes of beam-splitter output
$|\phi^{(20)}_{0}\,(\beta)\rangle$ versus beam-splitter angle $\beta$. $N$
= 20 photons.}
\label{EPRBS}
\end{figure}

\begin{figure}
\caption{Modulus of the quantum amplitudes of beam-splitter output
$|10\,m\,(\beta)\rangle$ versus beam-splitter angle $\beta$. $N$ = 20 
photons.}
\label{3DNNBS}
\end{figure}

\begin{figure}
\caption{Modulus (top) and phase (bottom) of the quantum amplitudes of
$|10\ 0(90^{\rm o})\rangle$ (left) and $|10\ 0(85.5^{\rm o})\rangle$
(right). The phase is $\pi/2$ for all nonzero components at $\rm 90^{o}$.}
\label{NN855BS}
\end{figure}

\begin{figure}
\caption{Modulus of the quantum amplitudes of $|j\,0\,(\beta_{Q})\rangle$
for $N$ = 200, 2000, and 20000 photons (top to bottom). All states have
almost identical appearance. See also $|10\,0\,(\beta_{Q})\rangle$ in
Fig.~\ref{NN855BS}.}
\label{NN895BS}
\end{figure}

\begin{figure}
\caption{Modulus of the quantum amplitudes of beam-splitter output
$(|j\,\frac{1}{2}\,(\beta)\rangle
+|j\,-\frac{1}{2}\,(\beta)\rangle)/\sqrt{2}$ versus beam-splitter angle
$\beta$. $N$ = 21 photons.}
\label{3D2pt}
\end{figure}

\begin{figure}
\caption{Teleportation fidelity using the quasi-EPR resource
$(|j\,1/2\,(\beta)\rangle + |j\,-1/2\,(\beta)\rangle)/\sqrt{2}$, versus
beam-splitter angle $\beta$ and number-sum measurement result $q$. $N$ = 21
photons.}
\label{F3D2pt}
\end{figure}

\begin{figure}
\caption{Teleportation fidelity using the quasi-EPR resource
$(|j\,0\,(\beta_{Q})\rangle$, versus beam-splitter angle $\beta$ for
number-sum measurement result $q=19$. $N$ = 20 photons.}
\label{Fj0}
\end{figure}

\end{document}